\documentclass[aps,showpacs,amsmath,amssymb,preprint,prb,floatfix,superscriptaddress,a4]{revtex4}

\usepackage[dvips]{graphicx}
\usepackage{amsmath}

\begin{document}

\title{Estimation of the spatial decoherence time in circular quantum dots}

\author{Michael Genkin}
\affiliation{Atomic Physics, Stockholm University, AlbaNova, S-10691 Stockholm, Sweden}

\author{Erik Waltersson}
\affiliation{Atomic Physics, Stockholm University, AlbaNova, S-10691 Stockholm, Sweden}

\author{Eva Lindroth}
\affiliation{Atomic Physics, Stockholm University, AlbaNova, S-10691 Stockholm, Sweden}

\date{\today}

\begin{abstract}
We propose a simple phenomenological model to estimate the spatial decoherence time in quantum dots. The dissipative phase space dynamics
is described in terms of the density matrix and the corresponding Wigner function, which are derived from a master equation with Lindblad operators 
linear in the canonical variables. The formalism was initially developed to describe diffusion and dissipation in deep inelastic heavy ion 
collisions, but also an application to quantum dots is possible.
It allows us to study the dependence of the decoherence rate on the dissipation strength, the temperature and an external magnetic field, which is 
demonstrated in illustrative calculations on a circular GaAs one-electron quantum dot.
\end{abstract}


\maketitle

\section{Introduction}
Decoherence processes in semiconductor quantum dots have attracted a lot of interest in the last years, not only 
due to their relevance for a quantum computer implementation~\cite{Loss_98} but also because they present an 
experimentally accessible system to study the decoherence process in general~\cite{Folk_01,Htoo_01,Vago_04,Sang_06,Bern_06}.
As demonstrated in several theoretical 
works~\cite{Khae_02,Khae_03,Golo_04,Voro_05,Jaca_05,Thor_05,Zhan_06,Yao_06,Bhak_06,Deng_06,Deng_06_E,Chou_07,Seme_07,Jian_08,Stan_08,Zhan_08,Wood_08,
Yang_08,Hern_08,Tu_08}, there are different processes that can lead to decoherence in a quantum dot, like interaction with optical and acoustic phonons 
or hyperfine interactions, in particular through electron spin coupling to a bath of nuclear spins. The processes occur on different time 
scales and are sensitive to external parameters like temperature or external magnetic fields. The coupling to the environment can be 
treated within the Born-Markov approximation~\cite{Voro_05,Chou_07,Stan_08}, but also effects beyond the Markovian limit can play a 
role~\cite{Thor_05,Bhak_06,Deng_06,Deng_06_E,Jian_08,Tu_08}. However, the vast majority of the processes studied so far focuses on spin decoherence, 
mainly because it is the spin of the electron(s) that makes a quantum computational application of quantum dots possible~\cite{Loss_98}. 
Nevertheless, also spatial 
decoherence which is arising from dissipative phase space dynamics in the canonically conjugated coordinates and momenta 
(see e.g.~\cite{Zure_03} for a detailed discussion), can 
possibly become relevant. As an example, one can think of a situation where the space- and spin part of a wave function are connected by the 
fermionic total asymmetry condition. Also in such a scheme as recently suggested in~\cite{Walt_09},
where electromagnetic transitions in solid state devices are used for controlled operations, phase space dynamics can be important since
it is not only the spin but also the total angular momentum that plays a crucial role.

In the present work, we aim to estimate the spatial decoherence time scale and to study its dependence on the coupling strength to the environment, 
on the temperature and on an external magnetic field. The latter plays a role since it determines the cyclotron frequency and hence also the effective 
confinement strength, which, together with the temperature, was shown to influence the asymptotic spatial decoherence in quadratic 
potentials~\cite{Isar_07}. In addition, the magnetic field also explicitly influences the time evolution of the system.

The study is carried out using an analytical model in the Markovian limit with linear Lindblad 
operators, which was initially developed to study diffusion and 
dissipation in heavy ion collisions~\cite{Gupt_84,Sand_87}. We show that, with an appropriate choice of the involved constants, the model can also be 
used to describe a two-dimensional quantum dot in a perpendicularly applied external magnetic field (section~\ref{mod}). The temperature dependence
is incorporated in the diffusion coefficients which significantly determine the time behavior of the density matrix~\cite{Palc_00}. 
The model has the advantage that it is rather general and allows us to include environmental effects without the necessity to explicitly compute the 
system-environment-interaction. The latter is instead taken into account by phenomenological constants that emerge from the Lindblad operators. This 
is sometimes referred to as the reduced dynamics approach. In practical calculations, it is then only necessary to find appropriate values for these 
constants, 
depending on the environmental effects one wishes to consider. In the illustrative calculations shown here, we relate these effects to the 
electron-phonon-interaction, but one could also consider other effects within basically the same model without loss of generality, provided the 
required input (as, e.g., electron-phonon scattering rates in our case) is known at least approximately.
In section~\ref{deco}, we discuss how the decoherence time scale can be determined from the model, using the general results derived in 
Ref.~\cite{Sand_87}. It is demonstrated in more detail in illustrative calculations on a circular GaAs one-electron quantum 
dot in section~\ref{calc}, followed by a discussion and concluding remarks in section~\ref{concl}.

\section{Description of the model}\label{mod}
The Hamiltonian of a one-electron quantum dot with a harmonic confinement (which is a very common choice, 
see e.g.~\cite{Maks_90,Reim_02,Walt_07,Roth_08})
in the $xy$-plane and exposed to an external magnetic field $B$ pointing in $z$-direction reads
\begin{equation}
H=\frac{{\bf p}^2}{2m^{*}}+\frac{1}{2}m^*\omega_0^2 r^2+\frac{e^2}{8m^*}B^2r^2+\frac{e}{2m^*}BL_z+g^*\mu_bBS_z.
\end{equation}
Here, $\omega_0$ denotes the confinement strength, $r=\sqrt{x^2+y^2}$ is the radial polar coordinate,
$L_z$ and $S_z$ denote the $z$-component of the orbital angular momentum operator and spin operator, respectively, 
$\mu_b$ is the Bohr magneton and $m^*,g^*$ are the effective mass and effective $g$-factor for the used semiconductor. 
By defining the cyclotron frequency $\omega_c=eB/m^*$ and the effective frequency $\omega=\sqrt{\omega_0^2+\omega_c^2/4}$, it can be written as
\begin{equation}\label{H}
H=H_0+H_s
\end{equation}
where
$H_s=g^*\mu_bBS_z$ is the spin part and
\begin{equation}\label{H_0}
H_0=\frac{{\bf p}^2}{2m^{*}}+\frac{1}{2}m^*\omega^2 r^2+\frac{e}{2m^*}BL_z.
\end{equation}
The latter expression is a general Hamiltonian of a two-dimensional harmonic oscillator exposed to a perpendicular magnetic field. The phase space 
dynamics of such a system when coupled to the environment was, for example, studied semiclassically in~\cite{Dodo_85}
by means of a Fokker-Planck equation. Also an explicit
inclusion of a heat bath in this Hamiltonian is possible, which leads to non-Markovian dynamics as recently demonstrated
in~\cite{Kala_07} for a nuclear system. 
Another possible approach is the influence functional method~\cite{Anas_00}, which was used to study the decoherence dynamics
of two  coupled harmonic oscillators in a general environment, with their potential minima being separated by a finite distance~\cite{Chou_08}.
Here, however, we will adopt a simpler phenomenological picture to describe the 
quantum dot coupling to an external environment, based on a Markovian master equation~\cite{Lind_76} which is known to be valid in the weak coupling 
limit~\cite{Karr_97}. In this framework, dissipation and decoherence are described by Lindblad operators, which is 
a rather common approach~\cite{Gall_96,Haba_98,Isar_99,Isar_02,Diet_04,Isar_07}. The formalism is based on earlier work of 
Gupta {\it et al}~\cite{Gupt_84} and Sandulescu {\it et al}~\cite{Sand_87} which is briefly introduced below.

Due to excitation of internal degrees of freedom (i.e. the nucleons) in heavy ion collisions, dissipation is a rather important issue in its quantum 
mechanical description. It is common to describe such a dynamics in terms of dimensionless coordinates of proton and neutron asymmetry, defined as
$q_1=(Z_1-Z_2)/(Z_1+Z_2),\,q_2=(A_1-A_2)/(A_1+A_2)$ where $Z_{1},Z_2$ and $A_{1},A_2$ are the charges and masses of the colliding nuclei.
A model 
to couple these coordinates was suggested in~\cite{Gupt_84}. Later, this model was generalized in~\cite{Sand_87}, where the complete description
of the dissipative dynamics was explicitly derived from the Markovian master equation for the density matrix $\rho$ given below,
\begin{equation}\label{master}
\frac{{\rm d}\rho}{{\rm d}
t}=-\frac{i}{\hbar}[H,\rho]+\frac{1}{2\hbar}\sum_{j}\left(\left[V_{j}\rho,V_{j}^{\dagger}\right]+
\left[V_{j},\rho V_{j}^{\dagger}\right]\right),
\end{equation}
where $V_{j}$ is a set of Lindblad operators, and the considered Hamiltonian had the following form:
\begin{equation}\label{Ham}
H=\sum_{k=1}^2\left(\frac{1}{2m_{k}}p_{k}^2+\frac{m_{k}\omega_{k}^2}{2}q_{k}^2\right)+
\frac{1}{2}\sum_{k_1,k_2=1}^2\mu_{k_1k_2}\left(p_{k_1}q_{k_2}+q_{k_2}p_{k_1}\right)+\nu_{12}q_1q_2+\kappa_{12}p_1p_2.
\end{equation}
Here, $p_1,p_2$ are the canonically conjugated momenta to the charge and mass asymmetry coordinates. The appearing coupling 
constants can be partly calculated from the nuclear liquid drop model or determined by fitting to experimental data. However, if these constants are 
chosen as $\nu_{12}=0,\,\kappa_{12}=0,\,\mu_{11}=0,\,\mu_{22}=0$ and
\begin{equation}\label{constants}
q_1=x,\,q_2=y,\,p_1=p_x,\,p_2=p_y,\,m_1=m^*=m_2,\,\omega_1=\omega=\omega_2,\,\mu_{21}=\frac{eB}{2m^*}=-\mu_{12},
\end{equation}
the Hamiltonian $H_0$ from Eq.~(\ref{H_0}) is reproduced exactly. Moreover, following the same choice of linear Lindblad operators as 
in~\cite{Sand_87},
\begin{equation}\label{LindOp}
V_{j}=\sum_{k=1}^2\left(a_{jk}p_{k}+b_{jk}q_{k}\right),\quad j=1,2,3,4
\end{equation}
where $a_{jk},b_{jk}$ are complex numbers, we see that the spin part $H_s$ of the full quantum dot Hamiltonian~(\ref{H})
commutes with the full Hamiltonian as 
well as with the Lindblad operators, so that the resulting equations of motion of the first and second moments in the canonical variables are 
unaffected. In other words, all results derived in~\cite{Sand_87} also remain valid in our case. Thus, we will omit the derivation, solution
and discussion of
the equations of motion here and only briefly quote the main results relevant for our study.
At this point, we would also like to mention that, since the spin motion decouples, no spin dephasing effects are present in the spatial 
decoherence studied here. The spin relaxation and dephasing times (often called $T_1$ and $T_2$) can also be studied within an equations-of-motion 
approach (see e.g. Ref.~\cite{Jian_08} and the references within), which, however, requires different models and is not considered in the present 
work. 

We use the abbreviations
\begin{equation}
{\bf m}(t)=(\langle x\rangle,\langle y\rangle,\langle p_x\rangle,\langle p_y\rangle)^{\rm T}
\end{equation}
for the time-dependent expectation values of the canonical phase space operators and
\begin{equation}
\sigma(t)=\left(\begin{array}{cccc}
\sigma_{xx} & \sigma_{xy} & \sigma_{xp_x} & \sigma_{xp_y}\\
\sigma_{yx} & \sigma_{yy} & \sigma_{yp_x} & \sigma_{yp_y}\\
\sigma_{p_xx} & \sigma_{p_xy} & \sigma_{p_xp_x} & \sigma_{p_xp_y}\\
\sigma_{p_yx} & \sigma_{p_yy} & \sigma_{p_yp_x} & \sigma_{p_yp_y}
\end{array}\right)
\end{equation}
for the time-dependent symmetric covariance matrix, where the elements are defined as 
\begin{equation}
\sigma_{AB}=\sigma_{BA}=\frac{1}{2}\langle AB+BA\rangle-\langle A\rangle\langle B\rangle
\end{equation}
for any two operators $A,B$.
The time evolution of the expectation values is given by 
\begin{equation}\label{expval}
{\bf m}(t)=\exp(tY){\bf m}(0),
\end{equation}
where ${\bf m}(0)$ denotes the initial expectation values and $Y$ is the time evolution matrix, which, after insertion of the constants given 
in~(\ref{constants}) into the general result from~\cite{Sand_87}, becomes
\begin{equation}\label{Ymatrix}
Y=\left(\begin{array}{cccc}
-\lambda_{11} & -\lambda_{12}-eB/(2m^*) & 1/m^* & -\alpha_{12} \\
-\lambda_{21}+eB/(2m^*) & -\lambda_{22} & \alpha_{12} & 1/m^* \\
-m^*\omega^2 & \beta_{12} & -\lambda_{11} & -\lambda_{21}-eB/(2m^*) \\
-\beta_{12} & -m^*\omega^2 & -\lambda_{12}+eB/2m & -\lambda_{22}
\end{array}\right).
\end{equation}
The phenomenological dissipation constants $\lambda_{kl},\alpha_{12}$ and $\beta_{12}$ emerge from the Lindblad operators~(\ref{LindOp}) 
and are explicitly given by
\begin{eqnarray}\nonumber
\alpha_{12}&=&-{\rm Im}\langle{\bf a_1},{\bf a_2}\rangle,\\
\beta_{12}&=&-{\rm Im}\langle{\bf b_1},{\bf b_2}\rangle,\\
\label{abbrev}
\nonumber
\lambda_{kl}&=&-{\rm Im}\langle{\bf a_k},{\bf b_l}\rangle,
\end{eqnarray}
where the vectors ${\bf a_k},{\bf b_l}$ are defined as (cf. Eq.~(\ref{LindOp}))
\begin{equation}
{\bf a_k}=(a_{1k},a_{2k},a_{3k},a_{4k})^{\rm T},\quad{\bf b_k}=(b_{1k},b_{2k},b_{3k},b_{4k})^{\rm T},
\end{equation}
with the scalar product
\begin{equation}
\langle{\bf x},{\bf y}\rangle=\sum_{i=1}^{4}x_i^{*}y_i.
\end{equation}
However, since we consider a circular one electron quantum dot where the dynamics is symmetric in $x$ and $y$, we set the off-diagonal 
dissipation constants to be zero here and in the following (i.e. $\lambda_{12}=0=\lambda_{21}$ and $\alpha_{12}=0=\beta_{12}$) and, furthermore, demand 
$\lambda_{11}=\lambda_{22}\equiv\lambda$, hereby restricting the dissipation strength to a single phenomenological parameter.
For the covariance matrix, the following time evolution is derived:
\begin{equation}\label{vars}
\sigma(t)=\exp(tY)(\sigma(0)-\sigma(\infty))(\exp(tY))^{\rm T}+\sigma(\infty).
\end{equation}
Here, $\sigma(0)$ is the initial covariance matrix and $\sigma(\infty)$ its asymptote. The latter can be determined from a set of diffusion 
coefficients, which are given by
\begin{eqnarray}\nonumber
D_{p_kp_l}=D_{p_lp_k}=\frac{\hbar}{2}{\rm Re}\langle{\bf b_k},{\bf b_l}\rangle,\\
D_{q_kq_l}=D_{q_lq_k}=\frac{\hbar}{2}{\rm Re}\langle{\bf a_k},{\bf a_l}\rangle,\\
\nonumber
D_{q_kp_l}=D_{p_lq_k}=-\frac{\hbar}{2}{\rm Re}\langle{\bf a_k},{\bf b_l}\rangle,
\end{eqnarray}
where the notation $q_1=x,q_2=y,p_1=p_x,p_2=p_y$ is used. They are connected to the asymptotic covariance matrix by the relation
\begin{equation}\label{lineq}
Y\sigma(\infty)+\sigma(\infty)Y^{\rm T}=-2D,
\end{equation}
where $D$ is the symmetric diffusion matrix
\begin{equation}\label{Dmatrix}
D=\left(\begin{array}{cccc}
D_{xx} & D_{xy} & D_{xp_x} & D_{xp_y}\\
D_{yx} & D_{yy} & D_{p_xy} & D_{yp_y}\\
D_{p_xx} & D_{p_xy} & D_{p_xp_x} & D_{p_xp_y}\\
D_{p_yx} & D_{p_yy} & D_{p_yp_x} & D_{p_yp_y}
\end{array}\right).
\end{equation}
The choice of the diffusion coefficients is, in general, a non-trivial issue, since 
there are several conditions that have to be obeyed in order to preserve the non-negativity of the density matrix and the uncertainity relation.
This will not be discussed here (see e.g.~\cite{Dekk_84,Sand2_87,Adam_99,Adam2_99,Palc_00} for more details). 
In the present work, we use a two-dimensional 
extension of the commonly used temperature-dependent coefficients of a harmonic oscillator without further mixing, such that the diffusion matrix
is diagonal:
\begin{equation}\label{dcoeff}
D_{xx}=\frac{\hbar\lambda}{2m^*\omega}\coth\left(\frac{\hbar\omega}{2kT}\right)=D_{yy},\quad 
D_{p_xp_x}=\frac{\hbar\lambda m^*\omega}{2}\coth\left(\frac{\hbar\omega}{2kT}\right)=D_{p_yp_y},
\end{equation}
and $D_{ij}=0\,{\rm otherwise}$,
where $T$ is the temperature and $k$ the Boltzmann constant.
From the given time evolution of the first and second moments, one can obtain the Wigner function $f_W$ of the system, which is the best possible
quantum mechanical analogon to a classical phase space density (although it is, in general, not positive everywhere and, therefore, cannot be 
interpreted as a true density). The latter was found by means of Weyl operators in~\cite{Sand_87}:
\begin{equation}\label{wf}
f_W(x,y,p_x,p_y,t)=\frac{1}{\sqrt{{\rm det}(2\pi\sigma(t))}}\exp\left(-\frac{1}{2}({\bf
\xi}-{\bf m}(t))^{\rm T}\sigma(t)^{-1}({\bf \xi}-{\bf m}(t))\right)
\end{equation}
where ${\bf \xi}=(x,y,p_x,p_y)^{\rm T}$ is the phase space vector.
This agrees with the result obtained in earlier work~\cite{Wang_45,Agar_71,Dodo_85}. 
In the following, we use this result to calculate the decoherence rate.

\section{Decoherence rate}\label{deco}
It is by far not trivial to give a general definition of 'decoherence'. 
Often, it is simply referred to as 'loss of coherence' in a quantum system or as entanglement of the latter with its envitonment. 
Technically, however, the degree of decoherence can be expressed through the density matrix of a quantum system, or, more precisely,
through the damping of its off-diagonal elements~\cite{Joos_03,Schl_07,Mori_90,Haba_98,Isar_07}. We shall adopt this definition in the following,
although it is also possible to study decoherence directly in terms of the Wigner function~\cite{Fold_03}. 
The density matrix is connected to the Wigner function by the transformation
\begin{eqnarray}
&&\langle x,y|\rho|x',y'\rangle(t)= \\
\nonumber
&&\int\int{\rm d} p_x\,{\rm d} 
p_y\,\exp\left(\frac{i}{\hbar}(p_x(x-x')+p_y(y-y'))\right)f_W((x+x')/2,(y+y')/2,p_x,p_y,t),
\end{eqnarray}
which can be carried out analytically for the Wigner function~(\ref{wf}). 
After evaluating the two-dimensional Gaussian integral and introducing new coordinates
\begin{eqnarray}\nonumber
\Sigma_x=\frac{x+x'}{2},\quad\Delta_x=x-x',\\
\Sigma_y=\frac{y+y'}{2},\quad\Delta_y=y-y',
\end{eqnarray}
we arrive at the following expression:
\begin{eqnarray}\nonumber
\langle x,y|\rho|x',y'\rangle(t)=N\exp\left[
-K_1(\Sigma_x-\langle x\rangle)^2-K_2(\Sigma_y-\langle y\rangle)^2-K_3(\Sigma_x-\langle x\rangle)(\Sigma_y-\langle y\rangle)-\right.\\
\nonumber
\left.K_4\Delta_x^2-K_5\Delta_y^2+K_6\Delta_x\Delta_y+iK_7(\Sigma_x-\langle x\rangle)\Delta_x+iK_8(\Sigma_y-\langle y\rangle)\Delta_y+\right.\\
\label{K110}\left.iK_9(\Sigma_x-\langle x\rangle)\Delta_y+iK_{10}(\Sigma_y-\langle y\rangle)\Delta_x
+\frac{i}{\hbar}(\langle p_x\rangle\Delta_x+\langle p_y\rangle\Delta_y)\right].
\label{rhoxy}
\end{eqnarray}
The normalization factor is given by
\begin{equation}
N=\sqrt{\frac{4\pi^2}{\rm{det}(2\pi\sigma(t))(c_{33}c_{34}-c_{34}^2)}},
\end{equation}
where $c_{ij}=c_{ji}$ denote the  elements of the
inverse of the covariance matrix $\sigma(t)^{-1}$. 
The explicit form of the time-dependent constants $K_1(t)..K_{10}(t)$ in terms of the matrix elements $c_{ij}(t)$ is given below:
\begin{eqnarray}\nonumber
K_1(t)&=&\frac{c_{11}}{2}+\frac{c_{13}c_{14}c_{34}-\frac{1}{2}(c_{13}^2c_{44}+c_{14}^2c_{33})}{c_{33}c_{44}-c_{34}^2},\\
\nonumber
K_2(t)&=&\frac{c_{22}}{2}+\frac{c_{23}c_{24}c_{34}-\frac{1}{2}(c_{23}^2c_{44}+c_{24}^2c_{33})}{c_{33}c_{44}-c_{34}^2},\\
\nonumber
K_3(t)&=&c_{12}+\frac{c_{34}(c_{13}c_{24}+c_{14}c_{23})-(c_{13}c_{23}c_{44}+c_{14}c_{24}c_{33})}{c_{33}c_{44}-c_{34}^2},\\
\nonumber
K_4(t)&=&\frac{c_{44}}{2\hbar^2(c_{33}c_{44}-c_{34}^2)},\\
\nonumber
K_5(t)&=&\frac{c_{33}}{2\hbar^2(c_{33}c_{44}-c_{34}^2)},\\
\nonumber
K_6(t)&=&\frac{c_{34}}{\hbar^2(c_{33}c_{44}-c_{34}^2)},\\
K_7(t)&=&\frac{c_{34}c_{14}-c_{44}c_{13}}{\hbar(c_{33}c_{44}-c_{34}^2)},\\
\nonumber
K_8(t)&=&\frac{c_{34}c_{23}-c_{24}c_{33}}{\hbar(c_{33}c_{44}-c_{34}^2)},\\
\nonumber
K_9(t)&=&\frac{c_{34}c_{13}-c_{14}c_{33}}{\hbar(c_{33}c_{44}-c_{34}^2)},\\
\nonumber
K_{10}(t)&=&\frac{c_{34}c_{24}-c_{23}c_{44}}{\hbar(c_{33}c_{44}-c_{34}^2)}.
\end{eqnarray}
Note that also the expectation values are functions of time, governed by~(\ref{expval}).
In the particular case of a circular one electron quantum dot, the dynamics is considerably simplified due to the symmetry
in $x$ and $y$, because the dispersions of the terms quadratic in $\Sigma_x,\Sigma_y$ that determine the 
damping of the diagonal elements of the density matrix
are identical ($K_1=K_2$). The same holds for the dispersions of the terms quadratic in $\Delta_x,\Delta_y$ that describe the damping of the
off-diagonal elements ($K_4=K_5$), which, following~\cite{Joos_03,Schl_07,Mori_90,Isar_07}, 
allows us to define a single decoherence parameter
\begin{equation}
\delta_D(t)=\frac{1}{2}\sqrt{\frac{K_1}{K_4}}=\frac{1}{2}\sqrt{\frac{K_2}{K_5}}.\\
\end{equation}
The definition is such that $\delta_D=1$ corresponds to a perfectly coherent state and $\delta_D=0$ implies that coherence is lost. In the next 
section, we investigate its time behaviour under the influence of the magnetic field, the temperature and dissipation. At this point, we would 
like to emphasize that not only the decoherence degree but also other purely quantum mechanical quantities can be extracted 
from the Wigner function in the case studied here. If the quantum dot is prepared in a coherent state, the Wigner function is positive everywhere, and 
hence it coincides with the classical phase space density~\cite{Frie_06}. For example, the quantum mechanical probability density $\rho(x,y,t)$
is directly obtained from the Wigner function by setting $x=x',\,y=y'$ (i.e. $\Sigma_x=x,\,\Sigma_y=y,\,\Delta_x=0=\Delta_y$) in Eq.~(\ref{rhoxy}).

\section{Illustrative calculations}\label{calc}
\begin{figure}
\includegraphics[width=\textwidth]{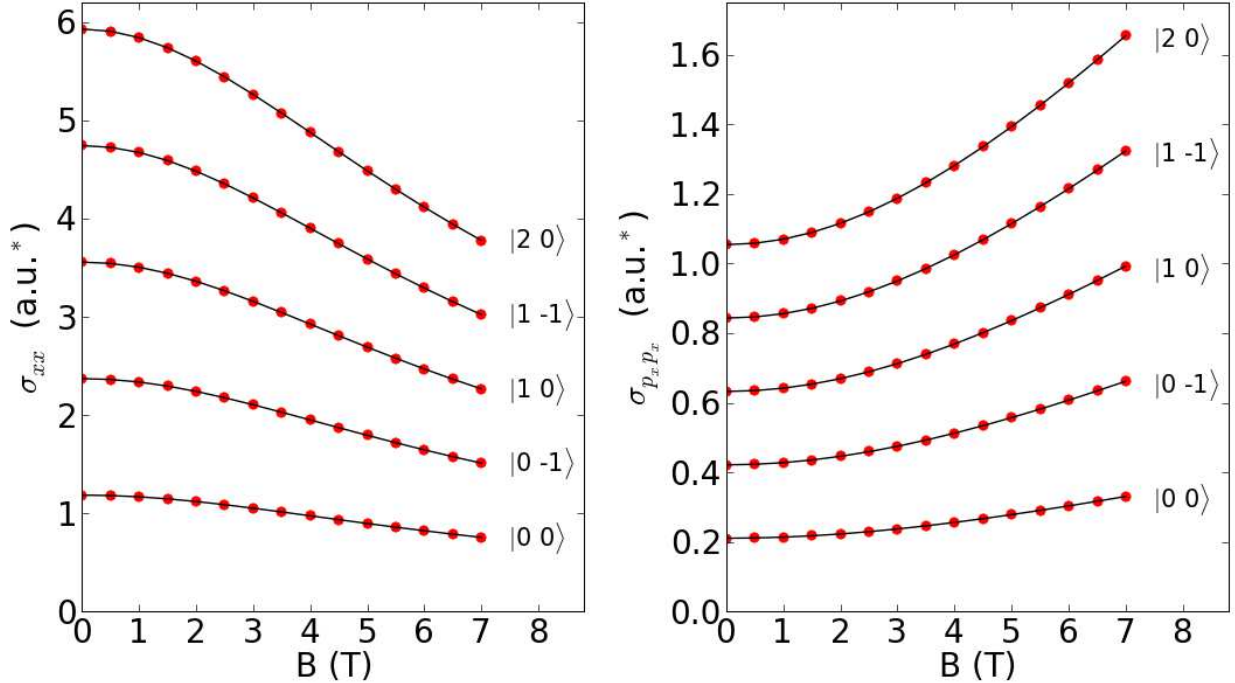}
 \caption{(Color online)
Variances in position and momentum shown for some one-electron GaAs quantum dot states $|nm\rangle$,
given in effective atomic units: $\sigma_{xx}$ has dimension length$^2$, where length is measured
in effective Bohr radii ($a_0^*\approx 9.794\,$nm) and $\sigma_{p_xp_x}$ is measured in Hartree$^*m^*$ where $m^*=0.067\,m_e$ and
Hartree$^*\approx 11.857\,$meV is the effective atomic energy unit.}\label{fig1}
\end{figure}

As an illustration, we consider a one-electron GaAs-quantum dot. Throughout this section, we use effective atomic units, i.e. atomic units
scaled with the GaAs material parameters $\epsilon_r=12.4$ (dielectric constant) and $m^*=0.067\,m_e$ where $m_e$ is the electron mass,
and a confinement strength of $\hbar\omega_0=5\,$meV.
Disregarding the spin part, which, as previously mentioned, is irrelevant for the phase space dynamics studied here, the solutions to the
Hamiltonian~(\ref{H_0}) can be written in terms of a principal and an angular quantum number
\begin{equation}\label{states}
\psi_{nm}(r,\varphi)=u_{nm}(r)e^{im\varphi}.
\end{equation}
Since we consider the system to be initially in a prepared state,
the initial conditions required for the time evolution of the first and second moments (Eqs.~\ref{expval} and \ref{vars}) 
are, unless given analytically,
calculated numerically for any state~(\ref{states}) using a B-Spline basis~\cite{Boor_78} as demonstrated in~\cite{Walt_07}. In particular, 
we have ${\bf m}(0)=0$ for the expectation values of $x,y,p_x,p_y$ and the initial covariance matrix has the following form:
\begin{equation}
\sigma(0)=\left(\begin{array}{cccc}
\sigma_{xx} & 0 & 0 & m\hbar \\
0 & \sigma_{yy} & -m\hbar & 0 \\
0 & -m\hbar & \sigma_{p_xp_x} & 0 \\
m\hbar & 0 & 0 & \sigma_{p_yp_y}
\end{array}\right)
\end{equation}
where $\sigma_{xx}=\sigma_{yy}$ and $\sigma_{p_{x}p_{x}}=\sigma_{p_{y}p_{y}}$ (the equality is again a consequence of the symmetry) are 
numerically calculated values. They depend on the effective confinement frequency and therefore
on the magnetic field since it influences the latter 
($\omega=\sqrt{\omega_0^2+\omega_c^2/4}$ where $\omega_c=eB/m^*$, cf. section~\ref{mod}). The explicit dependence is illustrated in 
figure~\ref{fig1}. 
This behavior can be understood qualitatively by considering a usual one-dimensional harmonic oscillator, where the variances of the $n$-th state are 
given by $\sigma_{xx}=(1+2n)\hbar/(2m^*\omega)$ and $\sigma_{p_xp_x}=(1+2n)\hbar m^*\omega/2$. Physically, this simply means that with increasing 
magnetic field the electron becomes more localized in position space and less localized in momentum space.
The uncertainty relation in each coordinate, however, 
does not depend on $\omega$ (and hence neither on the magnetic field) since it is determined by the product of the variances:
\begin{equation}\label{uncert} \sigma_{xx}\sigma_{p_xp_x}-\sigma_{xp_x}^2\geq\frac{1}{4}
\end{equation}
(and in the same way for $y,p_y$). Also, for the quantum dot states, the initial covarinaces $\sigma_{xp_x}$ and $\sigma_{yp_y}$ always 
vanish. Among all states, it is
only the ground state $|nm\rangle=|0\,0\rangle$ that has minimum uncertainty due to its ideal Gaussian shape,
while for excited states the Gaussian shape is disturbed and the uncertainty relation becomes a strict inequality. This corresponds to the
fact that the ground state of a harmonic oscillator is a particular case of a Glauber coherent state. Hence, $\delta_D(t=0)$ must be equal to unity
for the quantum dot ground state which agrees with the calculations and retrospectively confirms the imposed definition of the decoherence
parameter. 
\begin{figure}
\includegraphics[width=\textwidth]{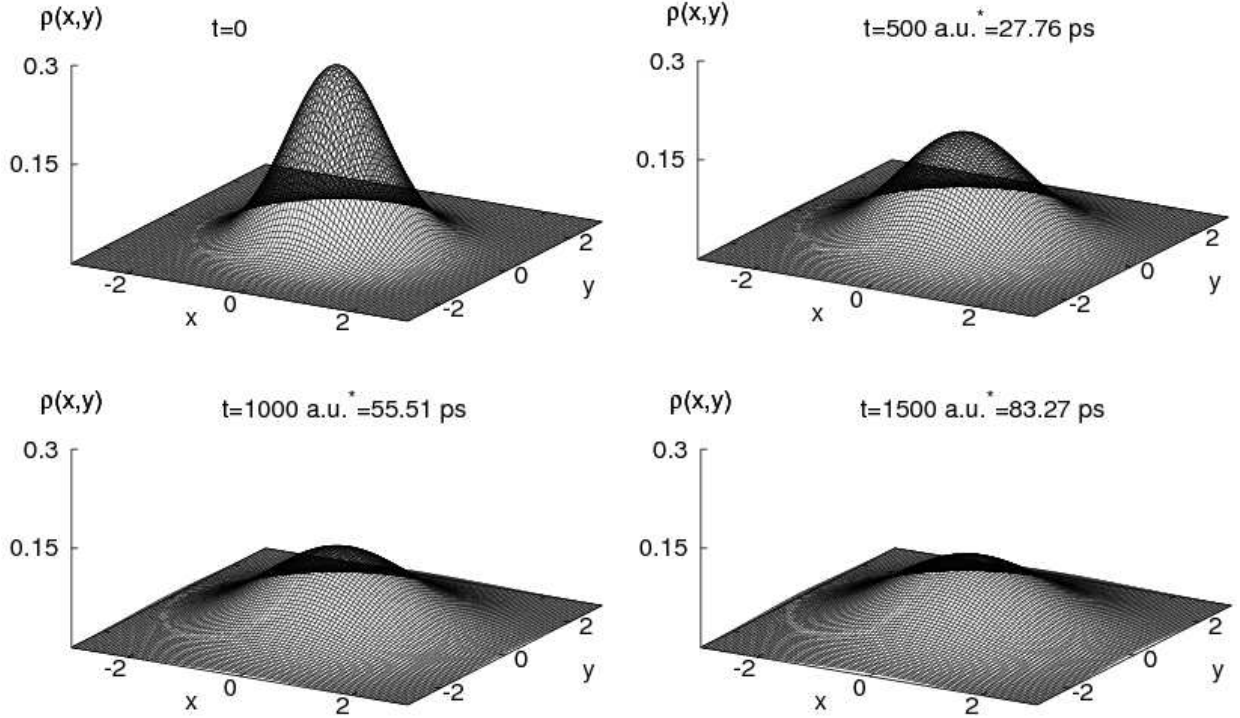}
\caption{Time evolution of the probability density in coordinate space shown for $B=1\,$T, $T=25\,$K and $\lambda=10^{-3}\omega$ where the dot is
initially prepared in the
ground state. The coordinates are given in effective Bohr radii $a_0^*\approx 9.794\,$nm and the probability density in units of $1/(a_0^*)^2$.}
\label{fig2}
\end{figure}

In addition to the previously discussed initial conditions, also the coupling strength to the environment needs to be determined. 
As already mentioned in the introduction,
the advantage of the model used here is that one can phenomenologically account for, in principle, any kind of environmental effects simply by choosing
appropriate values of the phenomenological constants. However, since the master equation~(\ref{master}) is only valid for weak coupling of the
reduced system to the environment, the ratio $\lambda/\omega$ should be much smaller than unity, which restricts the model to the description of
processes that obey this 
condition. Here, we consider the dissipative effects to be caused
by electron-phonon interactions and therefore the dissipation rate $\lambda$ is approximately given by the electron-phonon scattering rate. The latter 
ones were found to be typically of the order $\lambda\approx 10\,{\rm ns}^{-1}$ in GaAs
structures~\cite{Bock_94,Bert_05,Stav_05,Clim_06}. For the effective confinment strength chosen here, this yields $\lambda\approx 10^{-3}\omega$ and 
hence the Markovian approach can be seen as justified in the present study. Another aspect to be considered is that, apart from being incorporated in 
the diffusion coefficients~(Eq.~\ref{dcoeff}), the temperature also influences the phonon scattering rate. However, since the phonon scattering rates 
for different temperatures are not known exactly, the parameter $\lambda$ is varied independently in the relevant region at different temperatures to 
investigate its influence on the decoherence time scale. 

\begin{figure}
\includegraphics[width=\textwidth]{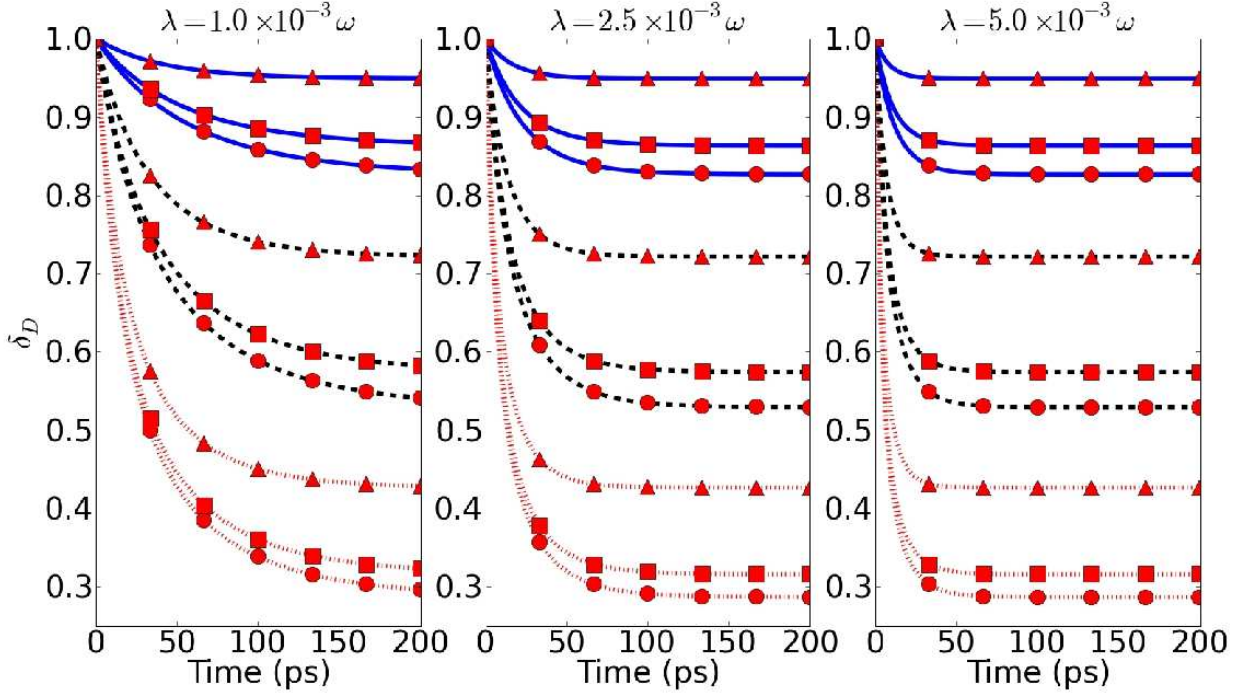}
 \caption{(Color online) Time evolution of the decoherence degree for three different environmental coupling strengths $\lambda$. The solid curves  
correspond to a temperature of $T=25\,$K, the dashed curves to $T=50\,$K and the dotted curves to $T=100\,$K, while the different magnetic fields are
indicated by symbols: $B=1\,$T (circles), $B=3\,$T (squares) and $B=7\,$T (triangles).
}\label{fig3}
\end{figure}

Figure~\ref{fig2} shows the time evolution of the probability density (obtained by setting $x=x',\,y=y'$ in Eq.~(\ref{rhoxy})) and figure~\ref{fig3}
the time evolution of the decoherence degree when the quantum dot is initially prepared in the ground state. 
We observe that the asymptotic degree of decoherence strongly
depends on the temperature and, more weakly, on the magnetic field strength. Naturally, higher temperatures lead to stronger decoherence, while
a strong magnetic field has a protective effect on quantum coherence. 
This corresponds to the result obtained for the one-dimensional case~\cite{Isar_07}, where the asymptotic decoherence degree was shown to be  
$\delta_D(\infty)=\tanh(\hbar\omega/(2kT))$ (Recall that the {\it effective} confinement $\omega$ increases with $B$). 
However, the time scale itself on which the system approaches its asymptotical 
decoherence is  determined by the coupling to the environment. 
Depending on the latter, decoherence occurs on a time scale between $20-200\,$ps. It is 
interesting to note that the rule of thumb to estimate the ratio of relaxation ($\tau_r$) and decoherence ($\tau_D$) time scale~\cite{Schl_07}
\begin{equation}\label{ratio}
\frac{\tau_r}{\tau_D}\approx\left(\frac{\sqrt{\sigma_{xx}}}{L_{dB}} \right)^2,
\end{equation}
where $L_{dB}$ is the thermal de Broglie wave length,
\begin{equation}
L_{dB}=\frac{\hbar}{\sqrt{2m^*kT}},
\end{equation}
holds in the case studied here, but the ratio is of the order of unity, which is a rather untypical behavior. In fact, in many situations decoherence 
is several orders of magnitude faster than dissipation and relaxation; For macroscopic systems, the ratio~(\ref{ratio}) becomes astronomically 
large~\cite{Isar_07,Schl_07,Zure_91}. In quantum dots, however, the wave packet spread can be of the same order as the thermal de 
Broglie wave length - a very remarkable property, once again displaying their fascinating features. 

\section{Conclusions}\label{concl}
Using a Markovian master equation approach with linear Lindblad operators, we investigated the dissipative phase space dynamics of a one-electron
quantum dot. We obtained the density matrix in coordinate representation from which an expression for the spatial decoherence parameter was derived.
With numerically calculated initial values for the first and second moments of a quantum dot in a prepared state, we analyzed the time evolution
of decoherence and the influence of the temperature and an external magnetic field. The phenomenological coupling to the environment was assumed to 
emerge from electron-phonon scattering. We found that the asymptotic decoherence strongly depends on the temperature and also on the magnetic field 
and the decoherence time scale is driven by environmental coupling strength, varying between a few and a few hundred picoseconds.

The model presented here has the advantage of being rather simple. The price one has to pay is, on the one hand, its phenomenological nature and,
on the other hand, the restriction to Markovian dynamics. Hence, the obtained results should be viewed with these limitations in mind. If, for example,
one aims to investigate decoherence arising from faster processes than electron-phonon scattering, the weak coupling limit is not 
necessarily valid, since the dissipation constant approaches the confinement strength. At the same time, the phenomenological nature also has the 
advantage that one is able to describe different kinds of interactions without loss of generality, just by choosing different values for the emerging 
constants, as long as the Markovian condition is not violated.

Further difficulties can occur, if, for example, the circular symmetry is disturbed or
anharmonic effects have to be accounted for. In that case, the off-diagonal coupling constants in Eqs.~(\ref{Ymatrix}) or~(\ref{Dmatrix}) 
may be non-zero, and it may also 
be quite non-trivial to find a single decoherence parameter in this case. It should be stressed that analytical solvability is provided for a purely 
harmonic confinement only, while for more complex potentials even a numerical solution is not always possible, since no finite system of the
equations of motions can be derived.

Another conclusion we can draw from the present study is that, if the temperature is very much smaller than the confinement
strength ($kT\ll\hbar\omega$), spatial decoherence should not be of significant relevance. Even at a temperature of $25\,$ K we see that
the asymptotic decoherence is about $\delta_D(t\rightarrow\infty)\approx 0.9$. However, in some experimental setups the temperatures are as low as
$T=100\,$mK, and hence we can conclude that spatial decoherence practically does not occur in that case.

As for further applications of the model presented here, it should be also suited to study tunneling processes in quantum dot molecules. 
Calculations on tunneling through one-dimensional parabolic potentials with dissipation described by linear Lindblad operators were, e.g.,
demonstrated in~\cite{Adam_98,Isar_00}, which can be extended to the two-dimensional case.

\section*{Acknowledgements}
Financial support from the G{\"o}ran Gustafsson Foundation and the Swedish science research council (VR) is gratefully acknowledged.


\begin{thebibliography}{66}
\expandafter\ifx\csname natexlab\endcsname\relax\def\natexlab#1{#1}\fi
\expandafter\ifx\csname bibnamefont\endcsname\relax
  \def\bibnamefont#1{#1}\fi
\expandafter\ifx\csname bibfnamefont\endcsname\relax
  \def\bibfnamefont#1{#1}\fi
\expandafter\ifx\csname citenamefont\endcsname\relax
  \def\citenamefont#1{#1}\fi
\expandafter\ifx\csname url\endcsname\relax
  \def\url#1{\texttt{#1}}\fi
\expandafter\ifx\csname urlprefix\endcsname\relax\def\urlprefix{URL }\fi
\providecommand{\bibinfo}[2]{#2}
\providecommand{\eprint}[2][]{\url{#2}}

\bibitem[{\citenamefont{Loss and Di{V}incenzo}(1998)}]{Loss_98}
\bibinfo{author}{\bibfnamefont{D.}~\bibnamefont{Loss}} \bibnamefont{and}
  \bibinfo{author}{\bibfnamefont{D.~P.} \bibnamefont{Di{V}incenzo}},
  \bibinfo{journal}{Phys. Rev. A} \textbf{\bibinfo{volume}{57}},
  \bibinfo{pages}{120} (\bibinfo{year}{1998}).

\bibitem[{\citenamefont{Folk et~al.}(2001)\citenamefont{Folk, Marcus, and
  Harris}}]{Folk_01}
\bibinfo{author}{\bibfnamefont{J.~A.} \bibnamefont{Folk}},
  \bibinfo{author}{\bibfnamefont{C.~M.} \bibnamefont{Marcus}},
  \bibnamefont{and} \bibinfo{author}{\bibfnamefont{J.~S.}
  \bibnamefont{Harris}}, \bibinfo{journal}{Phys. Rev. Lett.}
  \textbf{\bibinfo{volume}{87}}, \bibinfo{pages}{206802}
  (\bibinfo{year}{2001}).

\bibitem[{\citenamefont{Htoon et~al.}(2001)\citenamefont{Htoon, Kulik,
  Baklenov, Holmes, Takagahara, and Shih}}]{Htoo_01}
\bibinfo{author}{\bibfnamefont{H.}~\bibnamefont{Htoon}},
  \bibinfo{author}{\bibfnamefont{D.}~\bibnamefont{Kulik}},
  \bibinfo{author}{\bibfnamefont{O.}~\bibnamefont{Baklenov}},
  \bibinfo{author}{\bibfnamefont{A.~L.} \bibnamefont{Holmes}},
  \bibinfo{author}{\bibfnamefont{T.}~\bibnamefont{Takagahara}},
  \bibnamefont{and} \bibinfo{author}{\bibfnamefont{C.~K.} \bibnamefont{Shih}},
  \bibinfo{journal}{Phys. Rev. B} \textbf{\bibinfo{volume}{63}},
  \bibinfo{pages}{241303(R)} (\bibinfo{year}{2001}).

\bibitem[{\citenamefont{Vagov et~al.}(2004)\citenamefont{Vagov, Axt, Kuhn,
  Langbein, Borri, and Woggon}}]{Vago_04}
\bibinfo{author}{\bibfnamefont{A.}~\bibnamefont{Vagov}},
  \bibinfo{author}{\bibfnamefont{V.~M.} \bibnamefont{Axt}},
  \bibinfo{author}{\bibfnamefont{T.}~\bibnamefont{Kuhn}},
  \bibinfo{author}{\bibfnamefont{W.}~\bibnamefont{Langbein}},
  \bibinfo{author}{\bibfnamefont{P.}~\bibnamefont{Borri}}, \bibnamefont{and}
  \bibinfo{author}{\bibfnamefont{U.}~\bibnamefont{Woggon}},
  \bibinfo{journal}{Phys. Rev. B} \textbf{\bibinfo{volume}{70}},
  \bibinfo{pages}{201305(R)} (\bibinfo{year}{2004}).

\bibitem[{\citenamefont{Sanguinetti et~al.}(2006)\citenamefont{Sanguinetti,
  Poliani, Bonfanti, Guzzi, Grilli, Gurioli, and Koguchi}}]{Sang_06}
\bibinfo{author}{\bibfnamefont{S.}~\bibnamefont{Sanguinetti}},
  \bibinfo{author}{\bibfnamefont{E.}~\bibnamefont{Poliani}},
  \bibinfo{author}{\bibfnamefont{M.}~\bibnamefont{Bonfanti}},
  \bibinfo{author}{\bibfnamefont{M.}~\bibnamefont{Guzzi}},
  \bibinfo{author}{\bibfnamefont{E.}~\bibnamefont{Grilli}},
  \bibinfo{author}{\bibfnamefont{M.}~\bibnamefont{Gurioli}}, \bibnamefont{and}
  \bibinfo{author}{\bibfnamefont{N.}~\bibnamefont{Koguchi}},
  \bibinfo{journal}{Phys. Rev. B} \textbf{\bibinfo{volume}{73}},
  \bibinfo{pages}{125342} (\bibinfo{year}{2006}).

\bibitem[{\citenamefont{Bernardot et~al.}(2006)\citenamefont{Bernardot, Aubry,
  Tribollet, Testelin, Chamarro, Lombez, Braun, Marie, Amand, and
  G{\'e}rard}}]{Bern_06}
\bibinfo{author}{\bibfnamefont{F.}~\bibnamefont{Bernardot}},
  \bibinfo{author}{\bibfnamefont{E.}~\bibnamefont{Aubry}},
  \bibinfo{author}{\bibfnamefont{J.}~\bibnamefont{Tribollet}},
  \bibinfo{author}{\bibfnamefont{C.}~\bibnamefont{Testelin}},
  \bibinfo{author}{\bibfnamefont{M.}~\bibnamefont{Chamarro}},
  \bibinfo{author}{\bibfnamefont{L.}~\bibnamefont{Lombez}},
  \bibinfo{author}{\bibfnamefont{P.~F.} \bibnamefont{Braun}},
  \bibinfo{author}{\bibfnamefont{X.}~\bibnamefont{Marie}},
  \bibinfo{author}{\bibfnamefont{T.}~\bibnamefont{Amand}}, \bibnamefont{and}
  \bibinfo{author}{\bibfnamefont{J.-M.} \bibnamefont{G{\'e}rard}},
  \bibinfo{journal}{Phys. Rev. B} \textbf{\bibinfo{volume}{73}},
  \bibinfo{pages}{085301} (\bibinfo{year}{2006}).

\bibitem[{\citenamefont{Khaetskii et~al.}(2002)\citenamefont{Khaetskii, Loss,
  and Glazman}}]{Khae_02}
\bibinfo{author}{\bibfnamefont{A.~V.} \bibnamefont{Khaetskii}},
  \bibinfo{author}{\bibfnamefont{D.}~\bibnamefont{Loss}}, \bibnamefont{and}
  \bibinfo{author}{\bibfnamefont{L.}~\bibnamefont{Glazman}},
  \bibinfo{journal}{Phys. Rev. Lett.} \textbf{\bibinfo{volume}{88}},
  \bibinfo{pages}{186802} (\bibinfo{year}{2002}).

\bibitem[{\citenamefont{Khaetskii et~al.}(2003)\citenamefont{Khaetskii, Loss,
  and Glazman}}]{Khae_03}
\bibinfo{author}{\bibfnamefont{A.}~\bibnamefont{Khaetskii}},
  \bibinfo{author}{\bibfnamefont{D.}~\bibnamefont{Loss}}, \bibnamefont{and}
  \bibinfo{author}{\bibfnamefont{L.}~\bibnamefont{Glazman}},
  \bibinfo{journal}{Phys. Rev. B} \textbf{\bibinfo{volume}{67}},
  \bibinfo{pages}{195329} (\bibinfo{year}{2003}).

\bibitem[{\citenamefont{Golovach et~al.}(2004)\citenamefont{Golovach,
  Khaetskii, and Loss}}]{Golo_04}
\bibinfo{author}{\bibfnamefont{V.~N.} \bibnamefont{Golovach}},
  \bibinfo{author}{\bibfnamefont{A.}~\bibnamefont{Khaetskii}},
  \bibnamefont{and} \bibinfo{author}{\bibfnamefont{D.}~\bibnamefont{Loss}},
  \bibinfo{journal}{Phys. Rev. Lett.} \textbf{\bibinfo{volume}{93}},
  \bibinfo{pages}{016601} (\bibinfo{year}{2004}).

\bibitem[{\citenamefont{Vorojtsov et~al.}(2005)\citenamefont{Vorojtsov,
  Mucciolo, and Baranger}}]{Voro_05}
\bibinfo{author}{\bibfnamefont{S.}~\bibnamefont{Vorojtsov}},
  \bibinfo{author}{\bibfnamefont{E.~R.} \bibnamefont{Mucciolo}},
  \bibnamefont{and} \bibinfo{author}{\bibfnamefont{H.~U.}
  \bibnamefont{Baranger}}, \bibinfo{journal}{Phys. Rev. B}
  \textbf{\bibinfo{volume}{71}}, \bibinfo{pages}{205322}
  (\bibinfo{year}{2005}).

\bibitem[{\citenamefont{Jacak et~al.}(2005)\citenamefont{Jacak, Krasnyj, Jacak,
  Gonczarek, and Machnikowski}}]{Jaca_05}
\bibinfo{author}{\bibfnamefont{L.}~\bibnamefont{Jacak}},
  \bibinfo{author}{\bibfnamefont{J.}~\bibnamefont{Krasnyj}},
  \bibinfo{author}{\bibfnamefont{W.}~\bibnamefont{Jacak}},
  \bibinfo{author}{\bibfnamefont{R.}~\bibnamefont{Gonczarek}},
  \bibnamefont{and}
  \bibinfo{author}{\bibfnamefont{P.}~\bibnamefont{Machnikowski}},
  \bibinfo{journal}{Phys. Rev. B} \textbf{\bibinfo{volume}{72}},
  \bibinfo{pages}{245309} (\bibinfo{year}{2005}).

\bibitem[{\citenamefont{Thorwart et~al.}(2005)\citenamefont{Thorwart, Eckel,
  and Mucciolo}}]{Thor_05}
\bibinfo{author}{\bibfnamefont{M.}~\bibnamefont{Thorwart}},
  \bibinfo{author}{\bibfnamefont{J.}~\bibnamefont{Eckel}}, \bibnamefont{and}
  \bibinfo{author}{\bibfnamefont{E.~R.} \bibnamefont{Mucciolo}},
  \bibinfo{journal}{Phys. Rev. B} \textbf{\bibinfo{volume}{72}},
  \bibinfo{pages}{235320} (\bibinfo{year}{2005}).

\bibitem[{\citenamefont{Zhang et~al.}(2006)\citenamefont{Zhang, Dobrovitski,
  Al-Hassanieh, Dagotto, and Harmon}}]{Zhan_06}
\bibinfo{author}{\bibfnamefont{W.}~\bibnamefont{Zhang}},
  \bibinfo{author}{\bibfnamefont{V.~V.} \bibnamefont{Dobrovitski}},
  \bibinfo{author}{\bibfnamefont{K.~A.} \bibnamefont{Al-Hassanieh}},
  \bibinfo{author}{\bibfnamefont{E.}~\bibnamefont{Dagotto}}, \bibnamefont{and}
  \bibinfo{author}{\bibfnamefont{B.~N.} \bibnamefont{Harmon}},
  \bibinfo{journal}{Phys. Rev. B} \textbf{\bibinfo{volume}{74}},
  \bibinfo{pages}{205313} (\bibinfo{year}{2006}).

\bibitem[{\citenamefont{Yao et~al.}(2006)\citenamefont{Yao, Liu, and
  Sham}}]{Yao_06}
\bibinfo{author}{\bibfnamefont{W.}~\bibnamefont{Yao}},
  \bibinfo{author}{\bibfnamefont{R.~B.} \bibnamefont{Liu}}, \bibnamefont{and}
  \bibinfo{author}{\bibfnamefont{L.~J.} \bibnamefont{Sham}},
  \bibinfo{journal}{Phys. Rev. B} \textbf{\bibinfo{volume}{74}},
  \bibinfo{pages}{195301} (\bibinfo{year}{2006}).

\bibitem[{\citenamefont{BhaktavatsalaRao
  et~al.}(2006)\citenamefont{BhaktavatsalaRao, Ravishankar, and
  Subrahmanyam}}]{Bhak_06}
\bibinfo{author}{\bibfnamefont{D.~D.} \bibnamefont{BhaktavatsalaRao}},
  \bibinfo{author}{\bibfnamefont{V.}~\bibnamefont{Ravishankar}},
  \bibnamefont{and}
  \bibinfo{author}{\bibfnamefont{V.}~\bibnamefont{Subrahmanyam}},
  \bibinfo{journal}{Phys. Rev. A} \textbf{\bibinfo{volume}{74}},
  \bibinfo{pages}{022301} (\bibinfo{year}{2006}).

\bibitem[{\citenamefont{Deng and Hu}(2006{\natexlab{a}})}]{Deng_06}
\bibinfo{author}{\bibfnamefont{C.}~\bibnamefont{Deng}} \bibnamefont{and}
  \bibinfo{author}{\bibfnamefont{X.}~\bibnamefont{Hu}}, \bibinfo{journal}{Phys.
  Rev. B} \textbf{\bibinfo{volume}{73}}, \bibinfo{pages}{241303(R)}
  (\bibinfo{year}{2006}{\natexlab{a}}).

\bibitem[{\citenamefont{Deng and Hu}(2006{\natexlab{b}})}]{Deng_06_E}
\bibinfo{author}{\bibfnamefont{C.}~\bibnamefont{Deng}} \bibnamefont{and}
  \bibinfo{author}{\bibfnamefont{X.}~\bibnamefont{Hu}}, \bibinfo{journal}{Phys.
  Rev. B} \textbf{\bibinfo{volume}{74}}, \bibinfo{pages}{129902(E)}
  (\bibinfo{year}{2006}{\natexlab{b}}).

\bibitem[{\citenamefont{BenChouikha et~al.}(2007)\citenamefont{BenChouikha,
  Jaziri, and Bennaceur}}]{Chou_07}
\bibinfo{author}{\bibfnamefont{W.}~\bibnamefont{BenChouikha}},
  \bibinfo{author}{\bibfnamefont{S.}~\bibnamefont{Jaziri}}, \bibnamefont{and}
  \bibinfo{author}{\bibfnamefont{R.}~\bibnamefont{Bennaceur}},
  \bibinfo{journal}{Phys. Rev. A} \textbf{\bibinfo{volume}{76}},
  \bibinfo{pages}{062303} (\bibinfo{year}{2007}).

\bibitem[{\citenamefont{Semenov and Kim}(2007)}]{Seme_07}
\bibinfo{author}{\bibfnamefont{Y.~G.} \bibnamefont{Semenov}} \bibnamefont{and}
  \bibinfo{author}{\bibfnamefont{K.~W.} \bibnamefont{Kim}},
  \bibinfo{journal}{Phys. Rev. B} \textbf{\bibinfo{volume}{75}},
  \bibinfo{pages}{195342} (\bibinfo{year}{2007}).

\bibitem[{\citenamefont{Jiang et~al.}(2008)\citenamefont{Jiang, Wang, and
  Wu}}]{Jian_08}
\bibinfo{author}{\bibfnamefont{J.~H.} \bibnamefont{Jiang}},
  \bibinfo{author}{\bibfnamefont{Y.~Y.} \bibnamefont{Wang}}, \bibnamefont{and}
  \bibinfo{author}{\bibfnamefont{M.~W.} \bibnamefont{Wu}},
  \bibinfo{journal}{Phys. Rev. B} \textbf{\bibinfo{volume}{77}},
  \bibinfo{pages}{035323} (\bibinfo{year}{2008}).

\bibitem[{\citenamefont{Stano and Fabian}(2008)}]{Stan_08}
\bibinfo{author}{\bibfnamefont{P.}~\bibnamefont{Stano}} \bibnamefont{and}
  \bibinfo{author}{\bibfnamefont{J.}~\bibnamefont{Fabian}},
  \bibinfo{journal}{Phys. Rev. B} \textbf{\bibinfo{volume}{77}},
  \bibinfo{pages}{045310} (\bibinfo{year}{2008}).

\bibitem[{\citenamefont{Zhang et~al.}(2008)\citenamefont{Zhang, Konstantinidis,
  Dobrovitski, Harmon, Santos, and Viola}}]{Zhan_08}
\bibinfo{author}{\bibfnamefont{W.}~\bibnamefont{Zhang}},
  \bibinfo{author}{\bibfnamefont{N.~P.} \bibnamefont{Konstantinidis}},
  \bibinfo{author}{\bibfnamefont{V.~V.} \bibnamefont{Dobrovitski}},
  \bibinfo{author}{\bibfnamefont{B.~N.} \bibnamefont{Harmon}},
  \bibinfo{author}{\bibfnamefont{L.~F.} \bibnamefont{Santos}},
  \bibnamefont{and} \bibinfo{author}{\bibfnamefont{L.}~\bibnamefont{Viola}},
  \bibinfo{journal}{Phys. Rev. B} \textbf{\bibinfo{volume}{77}},
  \bibinfo{pages}{125336} (\bibinfo{year}{2008}).

\bibitem[{\citenamefont{Woods et~al.}(2008)\citenamefont{Woods, Reinecke, and
  Rajagopal}}]{Wood_08}
\bibinfo{author}{\bibfnamefont{L.~M.} \bibnamefont{Woods}},
  \bibinfo{author}{\bibfnamefont{T.~L.} \bibnamefont{Reinecke}},
  \bibnamefont{and} \bibinfo{author}{\bibfnamefont{A.~K.}
  \bibnamefont{Rajagopal}}, \bibinfo{journal}{Phys. Rev. B}
  \textbf{\bibinfo{volume}{77}}, \bibinfo{pages}{073313}
  (\bibinfo{year}{2008}).

\bibitem[{\citenamefont{Yang and Liu}(2008)}]{Yang_08}
\bibinfo{author}{\bibfnamefont{W.}~\bibnamefont{Yang}} \bibnamefont{and}
  \bibinfo{author}{\bibfnamefont{R.~B.} \bibnamefont{Liu}},
  \bibinfo{journal}{Phys. Rev. B} \textbf{\bibinfo{volume}{77}},
  \bibinfo{pages}{085302} (\bibinfo{year}{2008}).

\bibitem[{\citenamefont{Hernandez et~al.}(2008)\citenamefont{Hernandez,
  Greilich, Brito, Wiemann, Yakovlev, Reuter, Wieck, and Bayer}}]{Hern_08}
\bibinfo{author}{\bibfnamefont{F.~G.~G.} \bibnamefont{Hernandez}},
  \bibinfo{author}{\bibfnamefont{A.}~\bibnamefont{Greilich}},
  \bibinfo{author}{\bibfnamefont{F.}~\bibnamefont{Brito}},
  \bibinfo{author}{\bibfnamefont{M.}~\bibnamefont{Wiemann}},
  \bibinfo{author}{\bibfnamefont{D.~R.} \bibnamefont{Yakovlev}},
  \bibinfo{author}{\bibfnamefont{D.}~\bibnamefont{Reuter}},
  \bibinfo{author}{\bibfnamefont{A.~D.} \bibnamefont{Wieck}}, \bibnamefont{and}
  \bibinfo{author}{\bibfnamefont{M.}~\bibnamefont{Bayer}},
  \bibinfo{journal}{Phys. Rev. B} \textbf{\bibinfo{volume}{78}},
  \bibinfo{pages}{041303(R)} (\bibinfo{year}{2008}).

\bibitem[{\citenamefont{Tu and Zhang}(2008)}]{Tu_08}
\bibinfo{author}{\bibfnamefont{M.~W.~Y.} \bibnamefont{Tu}} \bibnamefont{and}
  \bibinfo{author}{\bibfnamefont{W.~M.} \bibnamefont{Zhang}},
  \bibinfo{journal}{Phys. Rev. B} \textbf{\bibinfo{volume}{78}},
  \bibinfo{pages}{235311} (\bibinfo{year}{2008}).

\bibitem[{\citenamefont{Zurek}(2003)}]{Zure_03}
\bibinfo{author}{\bibfnamefont{W.~H.} \bibnamefont{Zurek}},
  \bibinfo{journal}{Rev. Mod. Phys.} \textbf{\bibinfo{volume}{75}},
  \bibinfo{pages}{715} (\bibinfo{year}{2003}).

\bibitem[{\citenamefont{Waltersson et~al.}(2009)\citenamefont{Waltersson,
  Lindroth, Pilskog, and Hansen}}]{Walt_09}
\bibinfo{author}{\bibfnamefont{E.}~\bibnamefont{Waltersson}},
  \bibinfo{author}{\bibfnamefont{E.}~\bibnamefont{Lindroth}},
  \bibinfo{author}{\bibfnamefont{I.}~\bibnamefont{Pilskog}}, \bibnamefont{and}
  \bibinfo{author}{\bibfnamefont{J.~P.} \bibnamefont{Hansen}},
  \bibinfo{journal}{Phys. Rev. B} \textbf{\bibinfo{volume}{79}},
  \bibinfo{pages}{115318} (\bibinfo{year}{2009}).

\bibitem[{\citenamefont{Isar and Scheid}(2007)}]{Isar_07}
\bibinfo{author}{\bibfnamefont{A.}~\bibnamefont{Isar}} \bibnamefont{and}
  \bibinfo{author}{\bibfnamefont{W.}~\bibnamefont{Scheid}},
  \bibinfo{journal}{Physica A} \textbf{\bibinfo{volume}{373}},
  \bibinfo{pages}{298} (\bibinfo{year}{2007}).

\bibitem[{\citenamefont{Gupta et~al.}(1984)\citenamefont{Gupta, Muenchow,
  Sandulescu, and Scheid}}]{Gupt_84}
\bibinfo{author}{\bibfnamefont{R.~K.} \bibnamefont{Gupta}},
  \bibinfo{author}{\bibfnamefont{M.}~\bibnamefont{Muenchow}},
  \bibinfo{author}{\bibfnamefont{A.}~\bibnamefont{Sandulescu}},
  \bibnamefont{and} \bibinfo{author}{\bibfnamefont{W.}~\bibnamefont{Scheid}},
  \bibinfo{journal}{J. Phys. G: Nucl. Phys} \textbf{\bibinfo{volume}{10}},
  \bibinfo{pages}{209} (\bibinfo{year}{1984}).

\bibitem[{\citenamefont{Sandulescu et~al.}(1987)\citenamefont{Sandulescu,
  Scutaru, and Scheid}}]{Sand_87}
\bibinfo{author}{\bibfnamefont{A.}~\bibnamefont{Sandulescu}},
  \bibinfo{author}{\bibfnamefont{H.}~\bibnamefont{Scutaru}}, \bibnamefont{and}
  \bibinfo{author}{\bibfnamefont{W.}~\bibnamefont{Scheid}},
  \bibinfo{journal}{J. Phys. A: Math. Gen.} \textbf{\bibinfo{volume}{20}},
  \bibinfo{pages}{2121} (\bibinfo{year}{1987}).

\bibitem[{\citenamefont{Palchikov et~al.}(2000)\citenamefont{Palchikov,
  Adamian, Antonenko, and Scheid}}]{Palc_00}
\bibinfo{author}{\bibfnamefont{Y.~V.} \bibnamefont{Palchikov}},
  \bibinfo{author}{\bibfnamefont{G.~G.} \bibnamefont{Adamian}},
  \bibinfo{author}{\bibfnamefont{N.~V.} \bibnamefont{Antonenko}},
  \bibnamefont{and} \bibinfo{author}{\bibfnamefont{W.}~\bibnamefont{Scheid}},
  \bibinfo{journal}{J. Phys. A: Math. Gen.} \textbf{\bibinfo{volume}{33}},
  \bibinfo{pages}{4265} (\bibinfo{year}{2000}).

\bibitem[{\citenamefont{Maksym and Chakraborty}(1990)}]{Maks_90}
\bibinfo{author}{\bibfnamefont{P.~A.} \bibnamefont{Maksym}} \bibnamefont{and}
  \bibinfo{author}{\bibfnamefont{T.}~\bibnamefont{Chakraborty}},
  \bibinfo{journal}{Phys. Rev. Lett.} \textbf{\bibinfo{volume}{65}},
  \bibinfo{pages}{108} (\bibinfo{year}{1990}).

\bibitem[{\citenamefont{Reimann and Manninen}(2002)}]{Reim_02}
\bibinfo{author}{\bibfnamefont{S.~M.} \bibnamefont{Reimann}} \bibnamefont{and}
  \bibinfo{author}{\bibfnamefont{M.}~\bibnamefont{Manninen}},
  \bibinfo{journal}{Rev. Mod. Phys.} \textbf{\bibinfo{volume}{74}},
  \bibinfo{pages}{1283} (\bibinfo{year}{2002}).

\bibitem[{\citenamefont{Waltersson and Lindroth}(2007)}]{Walt_07}
\bibinfo{author}{\bibfnamefont{E.}~\bibnamefont{Waltersson}} \bibnamefont{and}
  \bibinfo{author}{\bibfnamefont{E.}~\bibnamefont{Lindroth}},
  \bibinfo{journal}{Phys. Rev. B} \textbf{\bibinfo{volume}{76}},
  \bibinfo{pages}{045314} (\bibinfo{year}{2007}).

\bibitem[{\citenamefont{Rothman and Mazziotti}(2008)}]{Roth_08}
\bibinfo{author}{\bibfnamefont{A.~E.} \bibnamefont{Rothman}} \bibnamefont{and}
  \bibinfo{author}{\bibfnamefont{D.~A.} \bibnamefont{Mazziotti}},
  \bibinfo{journal}{Phys. Rev. A} \textbf{\bibinfo{volume}{78}},
  \bibinfo{pages}{032510} (\bibinfo{year}{2008}).

\bibitem[{\citenamefont{Dodonov and Manko}(1985)}]{Dodo_85}
\bibinfo{author}{\bibfnamefont{V.~V.} \bibnamefont{Dodonov}} \bibnamefont{and}
  \bibinfo{author}{\bibfnamefont{O.~V.} \bibnamefont{Manko}},
  \bibinfo{journal}{Physica A} \textbf{\bibinfo{volume}{130}},
  \bibinfo{pages}{353} (\bibinfo{year}{1985}).

\bibitem[{\citenamefont{Kalandarov et~al.}(2007)\citenamefont{Kalandarov,
  Kanokov, Adamian, and Antonenko}}]{Kala_07}
\bibinfo{author}{\bibfnamefont{S.~A.} \bibnamefont{Kalandarov}},
  \bibinfo{author}{\bibfnamefont{Z.}~\bibnamefont{Kanokov}},
  \bibinfo{author}{\bibfnamefont{G.~G.} \bibnamefont{Adamian}},
  \bibnamefont{and} \bibinfo{author}{\bibfnamefont{N.~V.}
  \bibnamefont{Antonenko}}, \bibinfo{journal}{Phys. Rev. E}
  \textbf{\bibinfo{volume}{75}}, \bibinfo{pages}{031115}
  (\bibinfo{year}{2007}).

\bibitem[{\citenamefont{Anastopoulos and Hu}(2000)}]{Anas_00}
\bibinfo{author}{\bibfnamefont{C.}~\bibnamefont{Anastopoulos}}
  \bibnamefont{and} \bibinfo{author}{\bibfnamefont{B.~L.} \bibnamefont{Hu}},
  \bibinfo{journal}{Phys. Rev. A} \textbf{\bibinfo{volume}{62}},
  \bibinfo{pages}{033821} (\bibinfo{year}{2000}).

\bibitem[{\citenamefont{Chou et~al.}(2008)\citenamefont{Chou, Yu, and
  Hu}}]{Chou_08}
\bibinfo{author}{\bibfnamefont{C.~H.} \bibnamefont{Chou}},
  \bibinfo{author}{\bibfnamefont{T.}~\bibnamefont{Yu}}, \bibnamefont{and}
  \bibinfo{author}{\bibfnamefont{B.~L.} \bibnamefont{Hu}},
  \bibinfo{journal}{Phys. Rev. E} \textbf{\bibinfo{volume}{77}},
  \bibinfo{pages}{011112} (\bibinfo{year}{2008}).

\bibitem[{\citenamefont{Lindblad}(1976)}]{Lind_76}
\bibinfo{author}{\bibfnamefont{G.}~\bibnamefont{Lindblad}},
  \bibinfo{journal}{Commun. math. Phys.} \textbf{\bibinfo{volume}{48}},
  \bibinfo{pages}{119} (\bibinfo{year}{1976}).

\bibitem[{\citenamefont{Karrlein and Grabert}(1997)}]{Karr_97}
\bibinfo{author}{\bibfnamefont{R.}~\bibnamefont{Karrlein}} \bibnamefont{and}
  \bibinfo{author}{\bibfnamefont{H.}~\bibnamefont{Grabert}},
  \bibinfo{journal}{Phys. Rev. E} \textbf{\bibinfo{volume}{55}},
  \bibinfo{pages}{153} (\bibinfo{year}{1997}).

\bibitem[{\citenamefont{Gallis}(1996)}]{Gall_96}
\bibinfo{author}{\bibfnamefont{M.~R.} \bibnamefont{Gallis}},
  \bibinfo{journal}{Phys. Rev. A} \textbf{\bibinfo{volume}{53}},
  \bibinfo{pages}{655} (\bibinfo{year}{1996}).

\bibitem[{\citenamefont{Haba}(1998)}]{Haba_98}
\bibinfo{author}{\bibfnamefont{Z.}~\bibnamefont{Haba}}, \bibinfo{journal}{Phys.
  Rev. A} \textbf{\bibinfo{volume}{57}}, \bibinfo{pages}{4034}
  (\bibinfo{year}{1998}).

\bibitem[{\citenamefont{Isar et~al.}(1999)\citenamefont{Isar, Sandulescu, and
  Scheid}}]{Isar_99}
\bibinfo{author}{\bibfnamefont{A.}~\bibnamefont{Isar}},
  \bibinfo{author}{\bibfnamefont{A.}~\bibnamefont{Sandulescu}},
  \bibnamefont{and} \bibinfo{author}{\bibfnamefont{W.}~\bibnamefont{Scheid}},
  \bibinfo{journal}{Phys. Rev. E} \textbf{\bibinfo{volume}{60}},
  \bibinfo{pages}{6371} (\bibinfo{year}{1999}).

\bibitem[{\citenamefont{Isar and Scheid}(2002)}]{Isar_02}
\bibinfo{author}{\bibfnamefont{A.}~\bibnamefont{Isar}} \bibnamefont{and}
  \bibinfo{author}{\bibfnamefont{W.}~\bibnamefont{Scheid}},
  \bibinfo{journal}{Phys. Rev. A} \textbf{\bibinfo{volume}{66}},
  \bibinfo{pages}{042117} (\bibinfo{year}{2002}).

\bibitem[{\citenamefont{Dietz}(2004)}]{Diet_04}
\bibinfo{author}{\bibfnamefont{K.}~\bibnamefont{Dietz}}, \bibinfo{journal}{J.
  Phys. A: Math. Gen.} \textbf{\bibinfo{volume}{37}}, \bibinfo{pages}{6143}
  (\bibinfo{year}{2004}).

\bibitem[{\citenamefont{Dekker and Valsakumar}(1984)}]{Dekk_84}
\bibinfo{author}{\bibfnamefont{H.}~\bibnamefont{Dekker}} \bibnamefont{and}
  \bibinfo{author}{\bibfnamefont{M.~C.} \bibnamefont{Valsakumar}},
  \bibinfo{journal}{Phys. Lett. A} \textbf{\bibinfo{volume}{104}},
  \bibinfo{pages}{67} (\bibinfo{year}{1984}).

\bibitem[{\citenamefont{Sandulescu and Scutaru}(1987)}]{Sand2_87}
\bibinfo{author}{\bibfnamefont{A.}~\bibnamefont{Sandulescu}} \bibnamefont{and}
  \bibinfo{author}{\bibfnamefont{H.}~\bibnamefont{Scutaru}},
  \bibinfo{journal}{Ann. Phys.} \textbf{\bibinfo{volume}{173}},
  \bibinfo{pages}{277} (\bibinfo{year}{1987}).

\bibitem[{\citenamefont{Adamian
  et~al.}(1999{\natexlab{a}})\citenamefont{Adamian, Antonenko, and
  Scheid}}]{Adam_99}
\bibinfo{author}{\bibfnamefont{G.~G.} \bibnamefont{Adamian}},
  \bibinfo{author}{\bibfnamefont{N.~V.} \bibnamefont{Antonenko}},
  \bibnamefont{and} \bibinfo{author}{\bibfnamefont{W.}~\bibnamefont{Scheid}},
  \bibinfo{journal}{Nucl. Phys. A} \textbf{\bibinfo{volume}{645}},
  \bibinfo{pages}{376} (\bibinfo{year}{1999}{\natexlab{a}}).

\bibitem[{\citenamefont{Adamian
  et~al.}(1999{\natexlab{b}})\citenamefont{Adamian, Antonenko, and
  Scheid}}]{Adam2_99}
\bibinfo{author}{\bibfnamefont{G.~G.} \bibnamefont{Adamian}},
  \bibinfo{author}{\bibfnamefont{N.~V.} \bibnamefont{Antonenko}},
  \bibnamefont{and} \bibinfo{author}{\bibfnamefont{W.}~\bibnamefont{Scheid}},
  \bibinfo{journal}{Phys. Lett. A} \textbf{\bibinfo{volume}{260}},
  \bibinfo{pages}{39} (\bibinfo{year}{1999}{\natexlab{b}}).

\bibitem[{\citenamefont{Wang and Uhlenbeck}(1945)}]{Wang_45}
\bibinfo{author}{\bibfnamefont{M.~C.} \bibnamefont{Wang}} \bibnamefont{and}
  \bibinfo{author}{\bibfnamefont{G.~E.} \bibnamefont{Uhlenbeck}},
  \bibinfo{journal}{Rev. Mod. Phys.} \textbf{\bibinfo{volume}{17}},
  \bibinfo{pages}{323} (\bibinfo{year}{1945}).

\bibitem[{\citenamefont{Agarwal}(1971)}]{Agar_71}
\bibinfo{author}{\bibfnamefont{G.~S.} \bibnamefont{Agarwal}},
  \bibinfo{journal}{Phys. Rev. A} \textbf{\bibinfo{volume}{4}},
  \bibinfo{pages}{739} (\bibinfo{year}{1971}).

\bibitem[{\citenamefont{Joos et~al.}(2003)\citenamefont{Joos, Zeh, Kiefer,
  Giulini, Kupsch, and Stamatescu}}]{Joos_03}
\bibinfo{author}{\bibfnamefont{E.}~\bibnamefont{Joos}},
  \bibinfo{author}{\bibfnamefont{H.~D.} \bibnamefont{Zeh}},
  \bibinfo{author}{\bibfnamefont{C.}~\bibnamefont{Kiefer}},
  \bibinfo{author}{\bibfnamefont{D.}~\bibnamefont{Giulini}},
  \bibinfo{author}{\bibfnamefont{J.}~\bibnamefont{Kupsch}}, \bibnamefont{and}
  \bibinfo{author}{\bibfnamefont{I.~O.} \bibnamefont{Stamatescu}},
  \emph{\bibinfo{title}{Decoherence and the Appearance of a Classical World in
  Quantum Theory}} (\bibinfo{publisher}{Springer-Verlag},
  \bibinfo{year}{2003}).

\bibitem[{\citenamefont{Schlosshauer}(2007)}]{Schl_07}
\bibinfo{author}{\bibfnamefont{M.}~\bibnamefont{Schlosshauer}},
  \emph{\bibinfo{title}{Decoherence and the quantum-to-classical transition}}
  (\bibinfo{publisher}{Springer-Verlag}, \bibinfo{year}{2007}).

\bibitem[{\citenamefont{Morikawa}(1990)}]{Mori_90}
\bibinfo{author}{\bibfnamefont{M.}~\bibnamefont{Morikawa}},
  \bibinfo{journal}{Phys. Rev. D} \textbf{\bibinfo{volume}{42}},
  \bibinfo{pages}{2929} (\bibinfo{year}{1990}).

\bibitem[{\citenamefont{F{\"o}ldi et~al.}(2003)\citenamefont{F{\"o}ldi,
  Benedict, Czirj{\`a}k, and Moln{\`a}r}}]{Fold_03}
\bibinfo{author}{\bibfnamefont{P.}~\bibnamefont{F{\"o}ldi}},
  \bibinfo{author}{\bibfnamefont{M.~G.} \bibnamefont{Benedict}},
  \bibinfo{author}{\bibfnamefont{A.}~\bibnamefont{Czirj{\`a}k}},
  \bibnamefont{and}
  \bibinfo{author}{\bibfnamefont{B.}~\bibnamefont{Moln{\`a}r}},
  \bibinfo{journal}{Phys. Rev. A} \textbf{\bibinfo{volume}{67}},
  \bibinfo{pages}{032104} (\bibinfo{year}{2003}).

\bibitem[{\citenamefont{Friedrich}(2006)}]{Frie_06}
\bibinfo{author}{\bibfnamefont{H.}~\bibnamefont{Friedrich}},
  \emph{\bibinfo{title}{{T}heoretical {A}tomic {P}hysics}}
  (\bibinfo{publisher}{{S}pringer {V}erlag}, \bibinfo{year}{2006}).

\bibitem[{\citenamefont{deBoor}(1978)}]{Boor_78}
\bibinfo{author}{\bibfnamefont{C.}~\bibnamefont{deBoor}},
  \emph{\bibinfo{title}{A Practical Guide to Splines}}
  (\bibinfo{publisher}{Springer-Verlag}, \bibinfo{year}{1978}).

\bibitem[{\citenamefont{Bockelmann}(1994)}]{Bock_94}
\bibinfo{author}{\bibfnamefont{U.}~\bibnamefont{Bockelmann}},
  \bibinfo{journal}{Phys. Rev. B} \textbf{\bibinfo{volume}{50}},
  \bibinfo{pages}{17271} (\bibinfo{year}{1994}).

\bibitem[{\citenamefont{Bertoni et~al.}(2005)\citenamefont{Bertoni, Rontani,
  Goldoni, and Molinari}}]{Bert_05}
\bibinfo{author}{\bibfnamefont{A.}~\bibnamefont{Bertoni}},
  \bibinfo{author}{\bibfnamefont{M.}~\bibnamefont{Rontani}},
  \bibinfo{author}{\bibfnamefont{G.}~\bibnamefont{Goldoni}}, \bibnamefont{and}
  \bibinfo{author}{\bibfnamefont{E.}~\bibnamefont{Molinari}},
  \bibinfo{journal}{Phys. Rev. Lett.} \textbf{\bibinfo{volume}{95}},
  \bibinfo{pages}{066806} (\bibinfo{year}{2005}).

\bibitem[{\citenamefont{Stavrou and Hu}(2005)}]{Stav_05}
\bibinfo{author}{\bibfnamefont{V.~N.} \bibnamefont{Stavrou}} \bibnamefont{and}
  \bibinfo{author}{\bibfnamefont{X.}~\bibnamefont{Hu}}, \bibinfo{journal}{Phys.
  Rev. B} \textbf{\bibinfo{volume}{72}}, \bibinfo{pages}{075362}
  (\bibinfo{year}{2005}).

\bibitem[{\citenamefont{Climente et~al.}(2006)\citenamefont{Climente, Bertoni,
  Goldoni, and Molinari}}]{Clim_06}
\bibinfo{author}{\bibfnamefont{J.~I.} \bibnamefont{Climente}},
  \bibinfo{author}{\bibfnamefont{A.}~\bibnamefont{Bertoni}},
  \bibinfo{author}{\bibfnamefont{G.}~\bibnamefont{Goldoni}}, \bibnamefont{and}
  \bibinfo{author}{\bibfnamefont{E.}~\bibnamefont{Molinari}},
  \bibinfo{journal}{Phys. Rev. B} \textbf{\bibinfo{volume}{74}},
  \bibinfo{pages}{035313} (\bibinfo{year}{2006}).

\bibitem[{\citenamefont{Zurek}(1991)}]{Zure_91}
\bibinfo{author}{\bibfnamefont{W.~H.} \bibnamefont{Zurek}},
  \bibinfo{journal}{Phys. Today} \textbf{\bibinfo{volume}{44}},
  \bibinfo{pages}{36} (\bibinfo{year}{1991}).

\bibitem[{\citenamefont{Adamian et~al.}(1998)\citenamefont{Adamian, Antonenko,
  and Scheid}}]{Adam_98}
\bibinfo{author}{\bibfnamefont{G.~G.} \bibnamefont{Adamian}},
  \bibinfo{author}{\bibfnamefont{N.~V.} \bibnamefont{Antonenko}},
  \bibnamefont{and} \bibinfo{author}{\bibfnamefont{W.}~\bibnamefont{Scheid}},
  \bibinfo{journal}{Phys. Lett. A} \textbf{\bibinfo{volume}{244}},
  \bibinfo{pages}{482} (\bibinfo{year}{1998}).

\bibitem[{\citenamefont{Isar et~al.}(2000)\citenamefont{Isar, Sandulescu, and
  Scheid}}]{Isar_00}
\bibinfo{author}{\bibfnamefont{A.}~\bibnamefont{Isar}},
  \bibinfo{author}{\bibfnamefont{A.}~\bibnamefont{Sandulescu}},
  \bibnamefont{and} \bibinfo{author}{\bibfnamefont{W.}~\bibnamefont{Scheid}},
  \bibinfo{journal}{Eur. Phys. J. D} \textbf{\bibinfo{volume}{12}},
  \bibinfo{pages}{3} (\bibinfo{year}{2000}).

\end{thebibliography}
\end{document}